\documentclass[prl,twocolumn,superscriptaddress,floatfix]{revtex4}
\usepackage{graphicx}
\usepackage{nicefrac}
\usepackage{float}

\begin{document}

\title{Room temperature tetragonal noncollinear antiferromagnet Pt$_2$MnGa}
\author{S.~Singh} 
\affiliation{Max Planck Institute for Chemical Physics of Solids, N\"othnitzer Str.~40, D-01187 Dresden, Germany}
\author{S.~W.~D'Souza} 
\affiliation{Max Planck Institute for Chemical Physics of Solids, N\"othnitzer Str.~40, D-01187 Dresden, Germany}
\author{E.~Suard} 
\affiliation{Institut Laue-Langevin, BP 156, 38042 Grenoble Cedex~9,  France}
\author{L.~Chapon} 
\affiliation{Institut Laue-Langevin, BP 156, 38042 Grenoble Cedex~9, France}
\author{A.~Senyshyn} 
\affiliation{Forschungsneutronenquelle Heinz Maier-Leibnitz FRM-II, Technische Universit\"{a}t M\"{u}nchen, Lichtenbergstrasse 1, 85747 Garching,  Germany}
\author{V.~Petricek} 
\affiliation{Institute of Physics ASCR, Department of Structure Analysis, Na~Slovance~2, 18221 Praha, Czech Republic}
\author{Y.~Skourski} 
\affiliation{Dresden High Magnetic Field Laboratory (HLD-EMFL), Helmholtz-Zentrum Dresden-Rossendorf, D-01328 Dresden, Germany}
\author{M.~Nicklas} 
\affiliation{Max Planck Institute for Chemical Physics of Solids, N\"othnitzer Str.~40, D-01187 Dresden, Germany}
\author{C.~Felser} 
\affiliation{Max Planck Institute for Chemical Physics of Solids, N\"othnitzer Str.~40, D-01187 Dresden, Germany}
\author{S.~Chadov} 
\affiliation{Max Planck Institute for Chemical Physics of Solids, N\"othnitzer Str.~40, D-01187 Dresden, Germany}

\begin{abstract}
Here we present the tetragonal stoichiometric Heusler compound
Pt$_2$MnGa  with the noncollinear AFM order stable up to 350~K.
 It is resolved by the neutron diffraction as a helical spiral propagating along the tetragonal
axis. {\it Ab-initio} calculations suggest a pure exchange origin of the
spiral and explain its helical character being stabilized by a large basal
plane magnetocrystalline anisotropy (MCA). Together with the 
inversion-symmetric crystal structure, this provides a bi-stability
of a spiral with respect to the right- and left-handed magnetic helices.
Despite the large MCA, the long period of a helix might greatly facilitate
the switch of the helicity by the precessional reorientation,
  suggesting Pt$_2$MnGa as a  potential candidate for
 the vector-helicity based non-volatile magnetic memory.
\end{abstract}

\maketitle
\begin{figure}
\centering
\includegraphics[clip,width=0.7\linewidth]{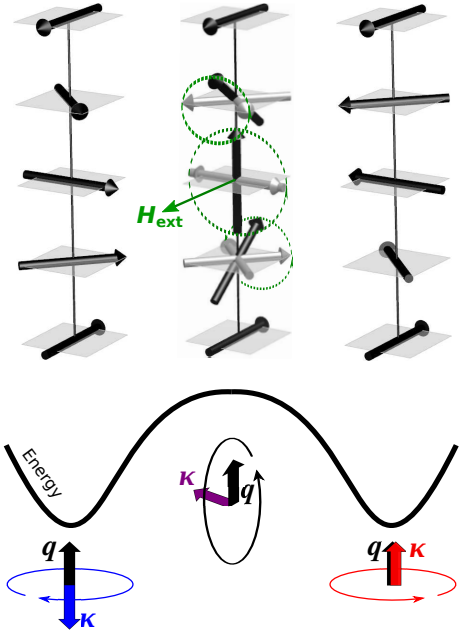}
\caption{
Application of the external magnetic pulse $\vec H_{\rm ext}$
perpendicular to the spiral wave vector $\vec q$ causes the precession of local moments (along the
dashed green circles). In case of the easy-plane MCA, the long period of a spiral greatly facilitates 
reorientation of helicity from ${\vec\kappa\uparrow\downarrow\vec q}$ (blue) to
${\vec\kappa\uparrow\uparrow\vec q}$ (red) or vice versa, since the top of the energy barrier between two stable screws is a cycloid
${\vec\kappa\perp\vec q}$ (magenta), in which only few atomic planes with magnetic moments orthogonal to $\vec H_{\rm ext}$ acquire the high energy.
\label{fig0}
}
\end{figure}
\begin{figure}
\centering
\includegraphics[clip,width=0.8\linewidth]{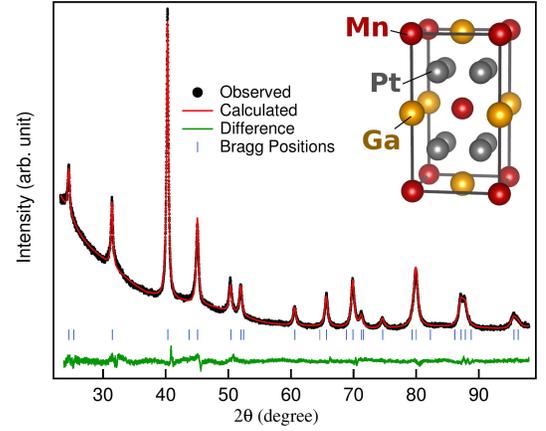}
\caption{{Crystal structure of Pt$_2$MnGa}. Rietveld refinement of
  the room-temperature XRD pattern assuming the tetragonal unit cell
  with $I4/mmm$ symmetry.  Observed and calculated  patterns, as well as their difference  are
  shown by black open circles, red and green solid lines,
  respectively. Blue vertical  ticks indicate the Bragg peak
  positions. The sketch of the unit cell is shown in the inset:
  red, yellow and gray spheres indicate  Mn, Ga and Pt atoms in $2a$, $2b$ and $4d$ Wyckoff
  positions, respectively.
\label{fig:str}}
\end{figure}

Antiferromagnets (AFMs) gain an increasing attention in the state-of-the-art applied and academic research.  
Their auxiliary role of a static support, enhancing the hardness of ferromagnetic
electrodes through the exchange bias effect in the conventional
microelectronics, has been broadly extended by the new prospectives in
spintronics applications. For instance, by studying the magnetoresistance effects typically exploited in
spintronics~\cite{PWM+11}, it has been demonstrated that Ir-Mn AFM, utilized as an active medium in a
tunneling magnetoresistance device, exhibits a 160\,\% tunneling anisotropic magnetoresistance at 4\,K  in weak magnetic fields
of 50\,mT or less. AFMs also facilitate the current-induced
switching of their order parameter~\cite{NDHM06,GL10,GL14} due to the absence of the shape anisotropy and the action of spin torques through the entire volume. 
For instance, a relatively low critical current density of
4.6\,MA/cm$^2$ was reported for the collinear AFM $\text{CuMnAs}$~\cite{WHZ+16}.
Additional nontrivial spintronic effects originating from a non-vanishing
Berry phase might occur in the noncollinear AFMs~\cite{Gom15}.
For instance, the noncollinear planar AFMs with the absence of mirror symmetry,
such as Mn$_3$Ir, where predicted to exhibit the anomalous Hall~\cite{CNM14,KF14}, Kerr and other effects characterized
by the same spatial tensor shape~\cite{SLWE15}, which where not encountered in AFM systems so far.\\
\indent Another set of specific properties, alternative to the above mentioned systems, are provided by
one-dimensional long-range AFM modulations, such as cycloidal - ${{\vec q}\perp({\vec e}_{i}\times\vec{e}_j)}$ and screw (or helical) - ${\vec q \parallel({\vec e}_{i}\times\vec{e}_j)}$,  with $\vec{e}_{i,j}$
being the spin directions on $i$ and $j$ neighboring atomic sites sitting along the spiral propagation vector ${\vec q}$.  
These systems possess a specific order parameter ${\vec\kappa_{ij}=\vec e_i\times\vec e_j}$, denoted as
``chirality'' or ``helicity''.  E.g., in cycloidal AFM insulators $\vec\kappa$ is coupled
to the polarization vector ${\vec P\sim\vec q\times\vec\kappa}$, by leading  to the first-order ferroelectic
effect~\cite{KNB05,KHJ+05,Mos06,SD06}. For the screw-spiral order (${\vec q\times\vec\kappa=0}$)
it becomes possible only upon the additional specific condition, namely, when the crystal structure remains invariant
under inversion and rotations around $\vec\kappa$, but non-invariant
under $180^\circ$ rotation of the $\vec\kappa$-axis~\cite{Ari07,JCK+12}.
The information transfer in cycloidal spirals along the one-dimensional
atomic chains with the fixed $\vec\kappa$, stabilized by the surface Dzyaloshinskii-Moriya
mechanism, was demonstrated by switching their phase with external
magnetic field~\cite{MMW+12}. Such scheme is inapplicable to the screw
spirals due to their energy degeneracy with respect to the $\vec\kappa$
reversal, even if they are deposited on a surface~\cite{BKPW14}. Similar to the situation with the ferroelectric effect,
to fix a helicity of the screw spiral would require additional
symmetry constrains on the crystal structure~\cite{GPS+13}.
Despite that the cycloidal order seems to be more ubiquitous for
the applications, the aforementioned degeneracy between the left- and right-handed magnetic screws in
crystals with inversion symmetry might be considered as an alternative advantage. In particular, it allows to
directly associate a bit of information with the helicity.  The switch
of $\vec\kappa$ can be realized e.g., by an external magnetic
pulse ${\vec H_{\rm ext}\perp\vec\kappa}$ which reorients the spins
precessionally (see Fig.~\ref{fig0}). 
In this case, both stable ${\pm\vec\kappa}$ states 
would be connected over the energy barrier representing the
cycloidal-ordered state. \\
\indent Here we report on a similar AFM spiral magnetic order in the tetragonal Pt$_2$MnGa Heusler 
system, revealed by the neutron diffraction.  The present first-principles analysis justifies the non-relativistic exchange origin of a spiral,
confirms its experimentally deduced wave-vector ${\vec q\approx(0,0,\nicefrac{1}{5})}$ in units of ${2\pi/c}$ and
 suggests the screw-type order  caused by a moderate hard axis
 (tetragonal $c$-axis) magnetocrystalline anisotropy (MCA). \\
\indent To our knowledge, there are no reliable experimental results on this material in the literature. The single old report on
Pt$_2$MnGa~\cite{HC71} briefly refers it as $L2_1$ AFM with ${T_{\rm N}=75}$\,K, but no further details are given. 
Later on, Pt$_2$MnGa has been studied by {\it ab-initio} assuming
ferromagnetic ordering and revealed  that the tetragonal phase is more stable~\cite{SGD+11,FLZ+15}. In~\cite{SGD+11} it was only  mentioned that in Pt$_x$Ni$_{2-x}$MnGa alloy series ``the AFM correlations
become stronger by increasing $x$''. In more recent {\it ab-initio}
study~\cite{RC16} the nearest neighbor AFM order was found to be noticeably
higher in energy compared to the ferromagnetic. 
To clear the actual crystal and magnetic structure we prepared a
polycrystalline Pt$_2$MnGa sample (details are given in the Supplementary). The room-temperature crystal
structure (Fig.~\ref{fig:str}) was deduced from the Rietveld refinement of the x-ray diffraction data. 
All Bragg reflections can be indexed by assuming the tetragonal space group ${I4/mmm}$. The refined lattice parameters are ${a=b=4.02}$\,\AA,
${c=7.24}$\,\AA; 
Pt occupies  $4d$~(0,\,$\nicefrac{1}{2}$,\,$\nicefrac{1}{4}$), while Mn
and Ga  - $2a$~(0,\,0,\,0) and $2b$~(0,\,0,\,$\nicefrac{1}{2}$) Wyckoff
sites, respectively (see the inset in Fig.~\ref{fig:str}).
\begin{figure}
\centering
\includegraphics[width=1\linewidth, clip]{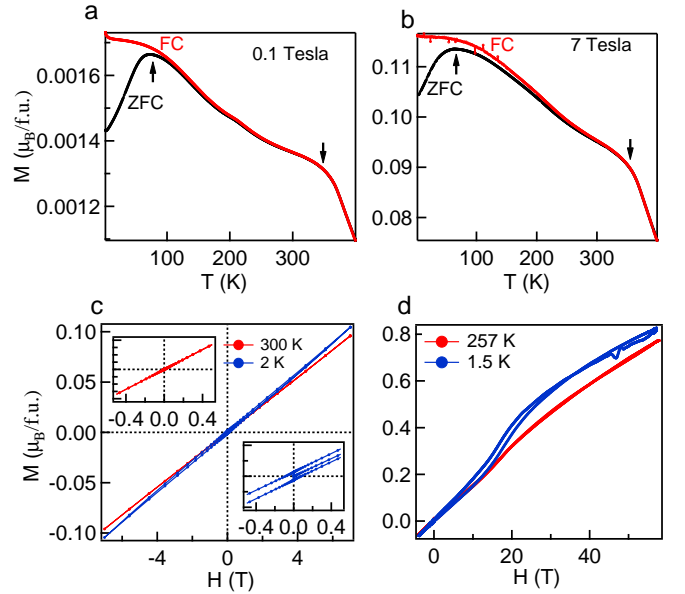}
\caption{{Magnetization of Pt$_2$MnGa as a function of temperature $T$
    and magnetic field $H$}: $M(T)$ at (a)~0.01\,T and (b)~7\,T; (c)~at
   300\,K and ~2\,K up to 7 T; $M(H)$ in the magnetic pulse of 60\,T at 257\,K and 1.5\,K (d). \label{fig:mag}}
\end{figure}\\
\indent The low-field $M(T)$ curves measured within the zero-field-cooled (ZFC) and field-cooled (FC) cycles are shown in
Fig.~\ref{fig:mag}\,a. The ZFC $M(T)$ shows a maximum at
${T\approx65}$\,K, absent in the FC regime. This observation typical for many Heusler alloys results from the polycrystalline configuration of the anisotropic crystallites.
The high-temperature behavior is similar in both ZFC and FC regimes and indicates the change of the
magnetic ordering at 350\,K.  
Overall, the amplitude of  $M(T)$ is very small in both ZFC and FC regimes, even at low  temperatures (2\,K) and higher fields (7\,T),
 (Fig.~\ref{fig:mag}\,b). Both isothermal $M(H)$ curves at 300\,K (Fig.~\ref{fig:mag}\,c) and 2\,K (Fig.~\ref{fig:mag}\,d) exhibit a non-saturating (straight line)
increase up to 7\,T, similar to the antiferromagnetic or paramagnetic
materials.  Only a narrow field hysteresis (Fig.~\ref{fig:mag}\,d, inset) indicates a very weak ferromagnetism at low temperature (2\,K).  
To probe the behavior of the system at very high magnetic fields, we
performed the 60\,T pulse measurements. Corresponding
$M(H)$ curves measured at 257 and 1.5\,K (Fig.~\ref{fig:mag}\,e, f) also
do not exhibit any saturation, by scaling almost
linearly with the field. Only at 1.5\,K a small hysteresis
within ${0<H<35}$\,T is observed indicating the  metamagnetic
transition around 14\,T induced by the non-equilibrium magnetic pulse.
All this clearly suggests that Pt$_2$MnGa is not a ferromagnet.\\
\indent To determine the actual magnetic order, the powder neutron
diffraction measurements were performed at 500\,K (above the magnetic
ordering), 300\,K and 3\,K. The crystal structure refinement of Pt$_2$MnGa of powder 
neutron diffraction data at 500\,K  gives similar results as of the room-temperature XRD, however, the substantial difference in the nuclear 
scattering amplitude (of 0.96, $-3.73$, and 7.29\,fm  for Pt, Mn, and
Ga, respectively) suggests a certain degree of disorder to be present in the actual atomic occupancies.
Fig.~\ref{fig:neut}\,a
\begin{figure*}
\centering
\includegraphics[width=1\textwidth,clip]{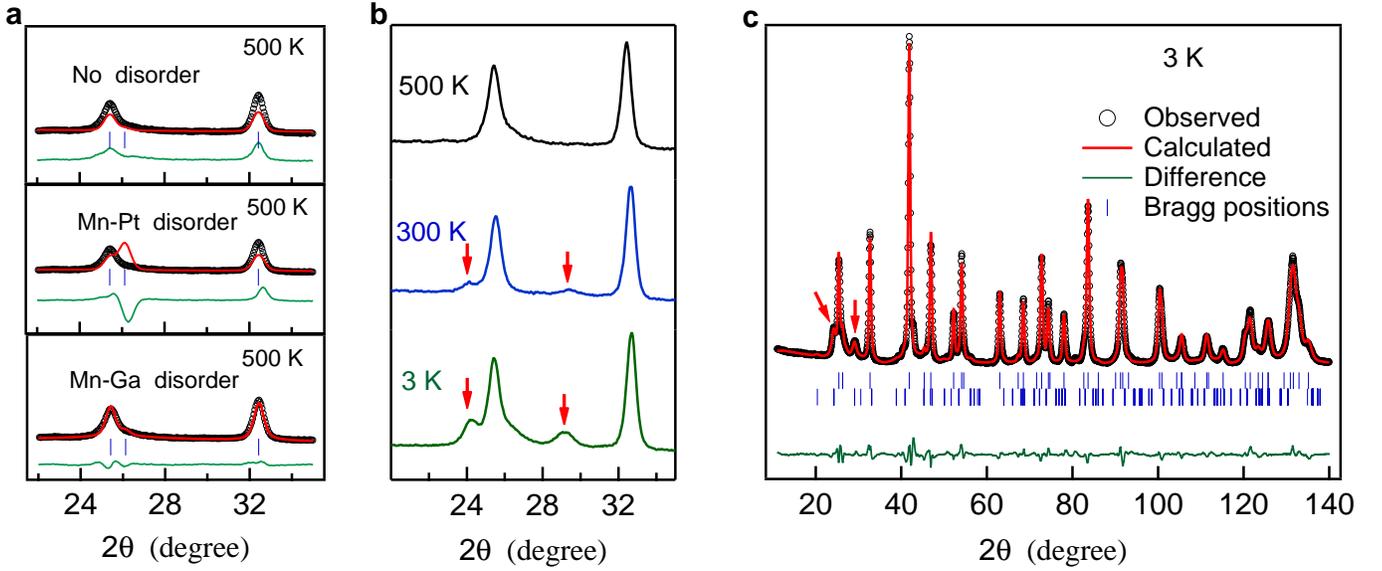}
\caption{Powder neutron diffraction studies on Pt$_2$MnGa. (a)~Rietveld refinements of 500\,K neutron diffraction pattern where the (002) and (110) Bragg peaks (black circles) have
been fitted (red solid lines) by assuming (i)~no disorder,
(ii)~Mn($2a$)/Pt($4d$) disorder, and (iii)~Mn($2a$)/Ga($2b$) disorder.  
The green curve shows the difference between observed and calculated
patterns. Vertical ticks are nuclear Bragg peak
positions. (b)~Comparison of neutron diffraction patterns at 500, 300 and
2\,K. Magnetic peaks are indicated by red arrows at 300 and 2\,K. (c)~The observed (black circle) and modeled (red solid line)
neutron diffraction pattern for Pt$_2$MnGa at 3\,K. The vertical arrows
indicate the magnetic peaks. Upper vertical ticks are nuclear Bragg
peak positions; lower vertical -  magnetic. \label{fig:neut}}
\end{figure*}
shows a comparison of the Rietveld refinement with the high-temperature
 spectrum  for (002) and (110) Bragg peaks, assuming several different 
 configurations. The chemically ordered model  gives a clear mismatch in
 the fit (upper panel of Fig.~\ref{fig:neut}\,a), the presence of random Mn($2a$)-Pt($4d$) disorder does not improve
it either. The most reasonable agreement gives a model with 33\,\%\ of
Mn($2a$)-Ga($2b$) disorder (lower panel in Fig.~\ref{fig:neut}\,a). The
refinement details for the  full range diffraction pattern at 500\,K 
is given in the Supplementary.\\
\indent By going to lower temperatures, we assumed the magnetic unit cell
with two Mn types in $2a$ and $2b$ sites having corresponding
occupancies. The Rietveld refinement of the magnetic phases observed at 300 and 3\,K was performed by accounting for 
the magnetic and atomic structures simultaneously. The comparison within
a narrow angular range ${20^\circ<2\theta<35^\circ}$ between 500, 300 and 3\,K neutron diffraction patterns is
given in Fig.~\ref{fig:neut}\,b. At 300 and 3\,K, the long-range magnetic
ordering is evidenced by the presence of two additional Bragg peaks, at ${2\theta\approx24.1^\circ}$ and $29.3^\circ$ (indicated by the red
arrows at 300 and 2\,K in Fig.~\ref{fig:neut}\,b), 
which closely corresponds to the commensurate reciprocal vector ${{\vec q}=(0,\,0,\,\nicefrac{1}{5})}$ in units of ${2\pi/c}$. Precise analysis reveals 
slightly incommensurate temperature-dependent variation: ${{\vec q}=(0,\,0,\,0.2066(1))}$ at 300\,K and ${(0,\,0,\,0.19(5))}$ at 3\,K. \\
\indent To specify more details of the magnetic ground state, we focus
on the 3\,K data exhibiting the highest intensity of the magnetic Bragg peaks. 
Its analysis suggests that the spiral order of the magnetic moments on
Mn atoms (in $2a$ and $2b$) would be more favorable, as it delivers the
magnetic moments amplitudes close to a reasonable Mn value of
4\,$\mu_{\rm B}$. In contrast, the collinear spin-wave
model leads to the values substantially exceeding 5\,$\mu_{\rm B}$. 
In the next step, we tried to distinguish which type of a spin spiral is more preferable. By assuming the spiral magnetic structure rotating
 in the $bc$-plane (cycloidal spiral) we obtained the moments of  4.33(13)\,$\mu_{\rm B}$ for both Mn in $2a$ and $2b$ sites. 
Although these values are acceptable, they slightly exceed those reported in the literature for Mn$(2a)$ in Mn-based
 Heusler alloys.  Finally, by assuming the spiral rotating within the $ab$-plane (screw spiral) leads to
 3.93(11)\,$\mu_{\rm B}$, a value which is somehow closer to those
 reported in  the literature. Since the rigorous experimental
 answer requires the polarized neutron spectroscopy data, in the following we will focus on  the first-principle analysis.\\
\indent To complete the experimental information we computed the {$\vec q$}-dependent total energies using the {\it ab-initio}
 LMTO method~\cite{PYA} adopting the local spin-density
approximation to the exchange-correlation~\cite{VWN80}. Since the method assumes
 perfectly ordered systems, we do not account for Mn/Ga chemical disorder
indicated by neutron scattering. The unit cell parameters were
taken from the present experimental refinement.\\
\indent First, we determined the preferential {$\vec q$} vector. Since the
parameters of the present magnetic modulation are defined mostly by the interplay of the isotropic exchange interactions which
have a largest energy scale (note, that the anisotropic
Dzyaloshinskii-Moriya interactions must be largely canceled by crystal symmetry),
it is practical to perform the non-relativistic calculations first. The
absence of the spin-orbit coupling allows to apply the generalized Bloch theorem and to study the spin
spirals within the chemical unit cell without going to the large
supercells. The energy dispersion computed along several symmetric
directions is shown in Fig.~\ref{fig:theory}\,a. 
\begin{figure*}
\centering
\includegraphics[width=1.0\linewidth,clip]{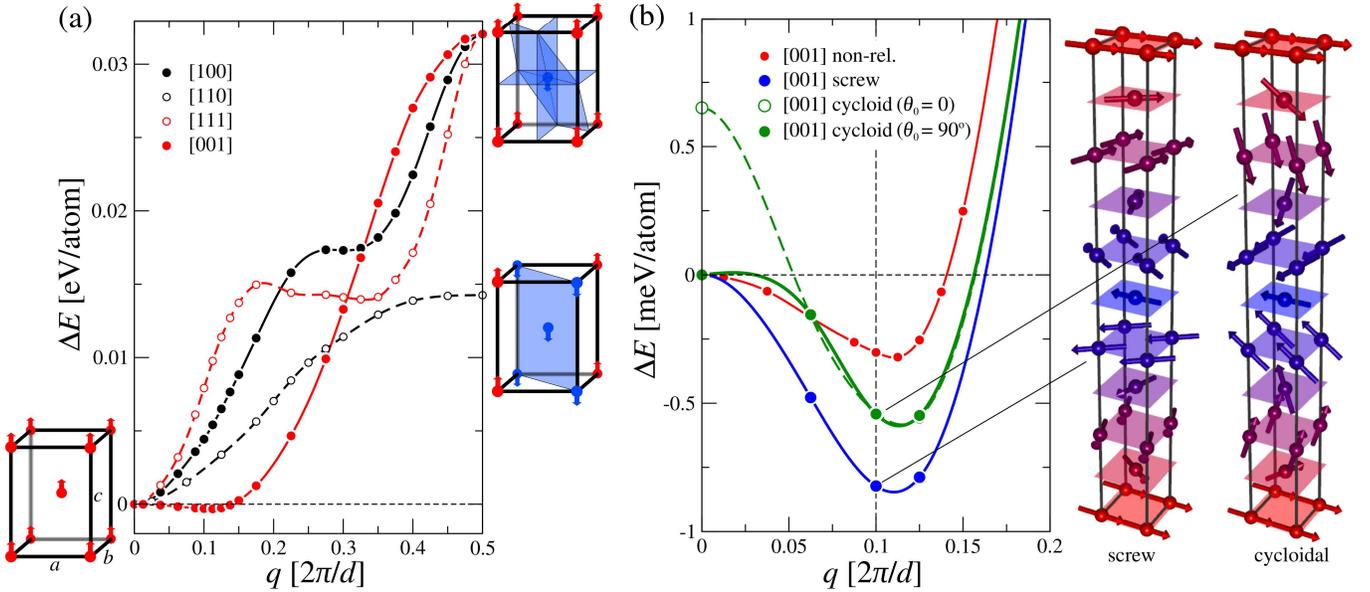}
\caption{Total energies ${\Delta E}$ calculated as functions of {$\vec q$} (in units of ${2\pi/d}$, where $d$ is the
  distance between the nearest Mn-containing planes orthogonal
  to $\vec q$\,). (a)~Non-relativistic regime: black solid and dashed lines
  refer to [100] (${d=a/2}$) and [110] (${d=a/\sqrt{2}}$) directions;
  red dashed and solid lines - to [111] (${d=(c/2)/\!\sqrt{2(c/a)^2+1}}$\,) and [001] (${d=c/2}$)
  directions, respectively. The energy zero is taken at ${q=0}$
  (ferromagnet). (b)~Detailed comparison of different regimes  along
  [001]: red, blue and green lines/points refer to the non-relativistic
  (same as [001] in (a)), screw- and cycloidal-type spirals, respectively. Cycloidal order has two
   variants: the first one (dashed green line, open circles) which represents at ${q=0}$ a hard-$c$-axis oriented
   ferromagnetic state (${\theta_0=0}$); the second one (solid green
   line, filled circles) represents at ${q=0}$ an easy-$ab$-plane oriented
   ferromagnetic state (${\theta_0=90^\circ}$). The energy zero corresponds to the easy-$ab$-plane
   ferromagnetic order. All lines are given for an eye guide. For several specific
  configurations the Mn($2a$) magnetic sublattice is shown explicitly; spin moments are colored from red to blue according to their phase.\label{fig:theory}}
\end{figure*}
In order to plot several curves along the same coordinate axis, we
give the $q$ length in the $2\pi/d$ units, where $d$ is the distance
between the nearest Mn-containing atomic planes orthogonal to ${\vec q}$. In this notation, ${q=0.5}$ always corresponds to the
antiparallel orientation of the spin moments in the nearest planes. 
As we see from Fig.~\ref{fig:theory}\,a, the energy dispersion in the $ab$-plane ([100] and [110] directions) is monotonous,
being characterized by a single minimum at ${q=0}$ (ferromagnet)
and a single maximum at ${q=0.5}$ (shortest AFM order). For the out-of-$ab$-plane directions one
observes a formation of a local minimum  within ${0<q<0.5}$. Whereas
along [111] direction it forms at rather high energy, along [001]
($c$-axis) it turns to global, supporting 
the experimental conclusions. Energy minimum vector is ${\vec q\approx(0,0,{0.11})=(0,0,0.22\cdot(2\pi/c))}$ (see also a more detailed plot in Fig.~\ref{fig:theory}\,b), which is very close to the
experimental one ${\vec q\approx(0,0,\nicefrac{1}{5}\cdot(2\pi/c))}$.
 Additional Monte-Carlo simulation of the classical Heisenberg model
parametrized by the {\it ab-initio} exchange coupling constants (see the
Supplementary) reasonably reproduces the magnetic ordering temperature
(${T_{\rm N}\approx350}$\,K) and reveal that the AFM order is set by
the interplay between the strong short-range parallel and the weaker long-range (7-th
shell) antiparallel interactions along the $c$-axis within the $2a$ sublattice. \\
\indent Next, we determine the type of the spin spiral (screw- or cycloidal),
which results from the relativistic effects. In this case the Bloch theorem does not hold and the magnetic order can be
studied only in supercells. Since in this case we can account only for the
commensurate modulations, the supercells must be sufficiently large 
to provide the energies at long wavelengths: a minimal supercell hosting a $(0,0,{\nicefrac{1}{5}\cdot\!(2\pi/c)})$
modulation contains at least five standard units. As we see
from Fig.~\ref{fig:theory}\,b, the relativistic effects substantially
deepen the spiral energy minimum and almost do not change the
corresponding $\vec q$ vector (${q\approx0.11}$). The $ab$-plane appears to be an
easy plane, since the screw-type spiral is more preferable in the whole range
of the wavelengths. The energy difference between the cycloidal and
the screw spirals is contributed by the MCA energy which has
a rather large magnitude for the Heusler class, being close to
0.65\,meV/atom (or 2.6\,meV/f.u.) at ${q=0}$. At the global 
minimum this energy difference is reduced more than twice. Due to the hard-$c$-axis
MCA, the cycloidal spiral can still be optimized in terms of a homogeneity (which must be distorted by the MCA), however
it will have a higher energy compared to the screw order anyway.
At the same time, as it is follows from the growing energy difference between
${\theta_0=0}$ and $90^\circ$ cases, the phase optimization of the
cycloidal spiral makes sense only for the small $q<0.06$, far from the global minimum. 
The amplitude of the Mn magnetic moment in the screw spiral has a tendency to grow by going from
${q=0}$ (ferromagnetic)  towards  ${q=0.5}$ (shortest AFM order),
though its absolute increase is relatively small: from about 3.7 to
3.8\,$\mu_{\rm B}$, which agrees with neutron data refinement.\\
\indent To conclude, we present a newly synthesized stoichiometric tetragonal
($I4/mmm$) Pt$_2$MnGa Heusler system exhibiting the room-temperature AFM spiral order
with the wave vector $\vec q=(0,0,\nicefrac{1}{5}\cdot2\pi/c)$, as it
follows from the neutron-diffraction refinement analysis. Certain degree
($\sim30$\,\%) of Mn-Ga chemical disorder was indicated.  
{\it Ab-initio} calculations (assuming the ordered system)
reasonably reproduce the experimental $\vec q$ vector indicating the exchange origin of the spiral. Monte-Carlo
simulations of the classical Heisenberg model parametrized with the
{\it ab-initio} exchange coupling constants reasonably reproduce the
Neel temperature and suggest the long-range antiparallel
Mn-Mn exchange (beyond the 6-th Mn shell) as a driving
mechanism for the AFM order. Relativistic calculations
indicate an  easy-$ab$-plane MCA, which stabilizes the screw (proper
screw, or fully helical) spiral type. Due to inversion symmetry, the left- and right-handed
spirals are stable and degenerate in energy. In spite of a large MCA,
the energy barrier between them can be efficiently overcome via the
precessional magnetization reorientation  induced by the magnetic pulse
perpendicular to the spiral axis. In this case, the barrier reduces to the energy difference between the screw and cycloidal spiral orders.
In particular, this suggests Pt$_2$MnGa as a convenient
candidate for the non-volatile magnetic memory based on the helicity vector as a bit of information. 

\begin{acknowledgments}
S.S. thanks to Alexander von Humboldt foundation for fellowship. 
S.C. thanks to A.N.~Yaresko (MPI-FKF Stuttgart) for providing his program code and discussions.
The work was financially supported by the ERC~AG~291472
``IDEA~Heusler!'' 
\end{acknowledgments}



\end{document}


\title {Supplementary information\\ Room temperature tetragonal noncollinear antiferromagnet Pt$_2$MnGa}

\author{S.~Singh} 
\affiliation{Max Planck Institute for Chemical Physics of Solids, N\"othnitzer Str.~40, D-01187 Dresden, Germany}
\author{S.~W.~D'Souza} 
\affiliation{Max Planck Institute for Chemical Physics of Solids, N\"othnitzer Str.~40, D-01187 Dresden, Germany}
\author{E.~Suard} 
\affiliation{Institut Laue-Langevin, BP 156, 38042 Grenoble Cedex~9,  France}
\author{L.~Chapon} 
\affiliation{Institut Laue-Langevin, BP 156, 38042 Grenoble Cedex~9, France}
\author{A.~Senyshyn} 
\affiliation{Forschungsneutronenquelle Heinz Maier-Leibnitz FRM-II, Technische Universit\"{a}t M\"{u}nchen, Lichtenbergstrasse 1, 85747 Garching, Germany}
\author{V.~Petricek} 
\affiliation{Institute of Physics ASCR, Department of Structure Analysis, Na~Slovance~2, 18221 Praha, Czech Republic}
\author{Y.~Skourski} 
\affiliation{Dresden High Magnetic Field Laboratory (HLD-EMFL), Helmholtz-Zentrum Dresden-Rossendorf, D-01328 Dresden, Germany}
\author{M.~Nicklas} 
\affiliation{Max Planck Institute for Chemical Physics of Solids, N\"othnitzer Str.~40, D-01187 Dresden, Germany}
\author{C.~Felser} 
\affiliation{Max Planck Institute for Chemical Physics of Solids, N\"othnitzer Str.~40, D-01187 Dresden, Germany}
\author{S.~Chadov} 
\affiliation{Max Planck Institute for Chemical Physics of Solids, N\"othnitzer Str.~40, D-01187 Dresden, Germany}

\begin{abstract}

\end{abstract}
\maketitle
\section{Experimental  details}

Polycrystalline ingot of Pt$_2$MnGa was prepared by melting appropriate quantities of  Pt, Mn and Ga of 99.99\% purity in an arc furnace. Ingots were then annealed at  1273\,K for 5 days to obtain homogeneity and subsequently quenched into ice water. Powder x-ray diffraction (XRD) at room temperature (RT)was done to investigate the sample quality and homogeneity using CuK$\alpha$ radiation. The composition of the sample
was confirmed using energy dispersive analysis of x-rays (EDAX) analysis which gave a composition Pt$_{50.98}$Mn$_{25.05}$Ga$_{23.98}$ (Pt$_{2.04}$Mn$_{1.0}$Ga$_{0.96}$), which we  refer to as Pt$_2$MnGa henceforth in the manuscript.
~The temeprature dependent ($M(T)$) and field dependent ($M(H)$) magnetization measurements were done using SQUID-VSM magnetometer.  The low-field $M(T)$ curves measured within the zero-field-cooled (ZFC) and field-cooled (FC) cycles  For the ZFC-measurement, the sample was cooled in zero field down to 2\,K and after 0.1\,T field was applied and then the data were recorded in heating cycle up to 400\,K. Subsequently,
the data were recorded in the same field (0.1\,T) by cooling from 400 down to 2\,K (FC).The magnetic isotherms, M(H), at 1.2\,K and 257 K have been done in a pulsed fields up to 60\,T. The pulsed magnetic field experiments were performed at the Dresden High Magnetic Field Laboratory. Neutron diffraction measurements were done at the D2B high-resolution neutron powder diffractometer (ILL, Grenoble). A vanadium cylinder was used as sample holder. The data were collected at 500\,K, 300\,K and 3\,K~ using a neutron wavelength of 1.59 \AA~ in the high-intensity mode. The analysis of diffraction patterns were done with Fullprof software package\cite{FullProf00}. 

\subsection{Nuclear structure from neutron diffraction}

\begin{figure}
\centering
\includegraphics[width=1\linewidth,clip]{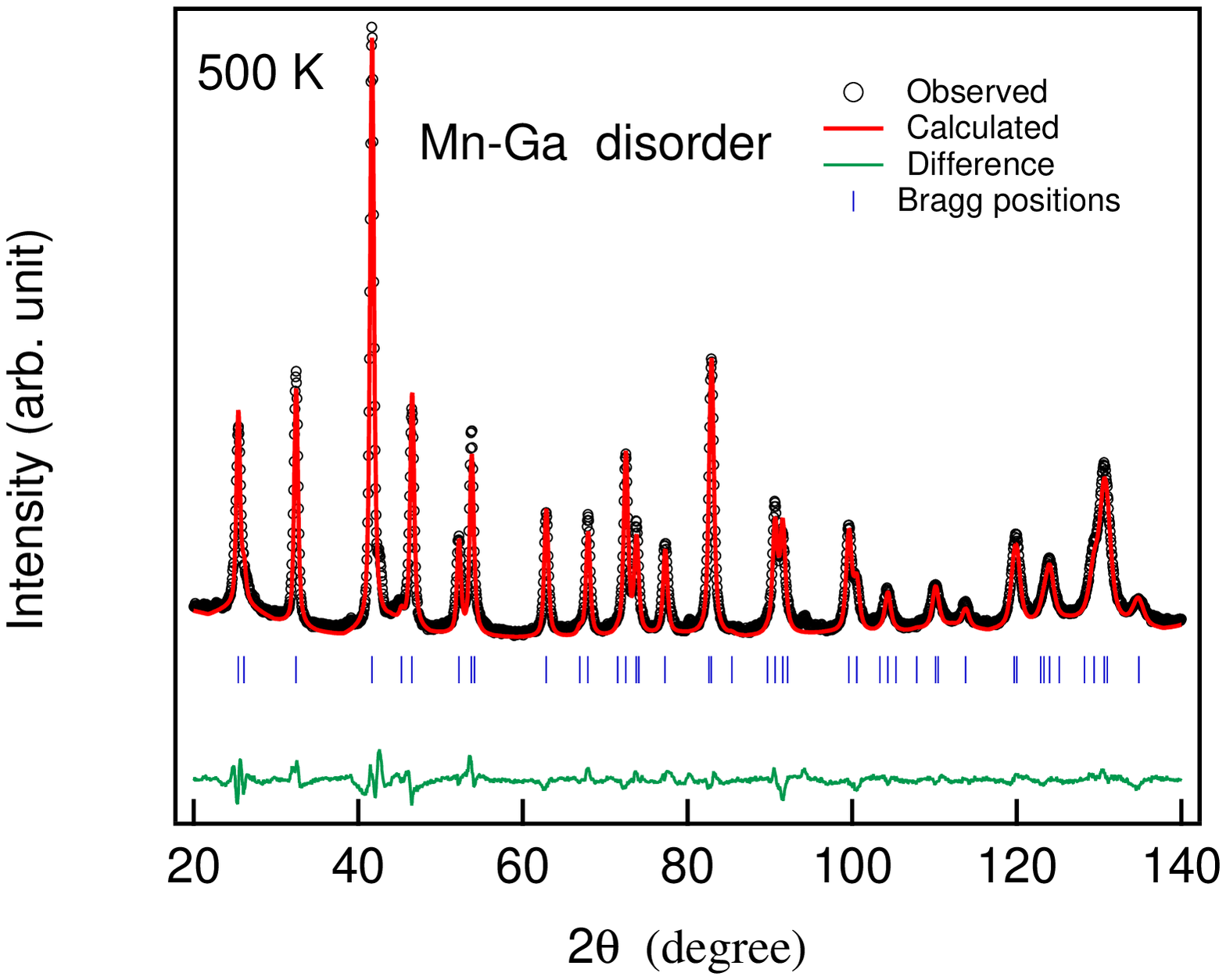}
\caption{ Rietveld refinement of powder neutron diffraction pattern of 500\,K in the $2\theta$ range of 20-100$^{\circ}$. The observed, calculated and difference patterns are shown by black,red  and green solid lines, respectively. The vertical ticks indicate the nuclear Bragg peak positions. 
\label{fig:neuc}}
\end{figure}

Fig.~\ref{fig:neuc} shows the observed and calculated neutron diffraction pattern of at 500\,K (paramagnetic phase) in the $2\theta$ range of 20-100$^{\circ}$. All the neutron diffraction  peaks can be indexed well by the tetragonal unit cell with refined lattice parameters are ${a=b=4.03}$\,\AA, ${c=7.23}$\,\AA. The Rietveld refinement was performed using space group $I4/mmm$ as in case of x-ray diffraction. In order case the Pt occupy the 4$d$ (0,\,0.5,\,0.25) position, while Mn and Ga occupied at  2$a$ (0,\,0,\,0) and 2$b$ (0,\,0,\,0.5) Wyckoff positions, respectively.  However, the  Rietveld analsyis shows that a substantial (~33\%) Mn(2$a$)-Ga(2$b$) antisite disorder exist in the sample. Therefore,33\%  Mn occupied at Ga (2$b$) site and similarly 30\% Ga occupied at Mn 2$a$ site.

\section{Theoretical details}

\begin{figure}
\centering
\includegraphics[width=1.0\linewidth,clip]{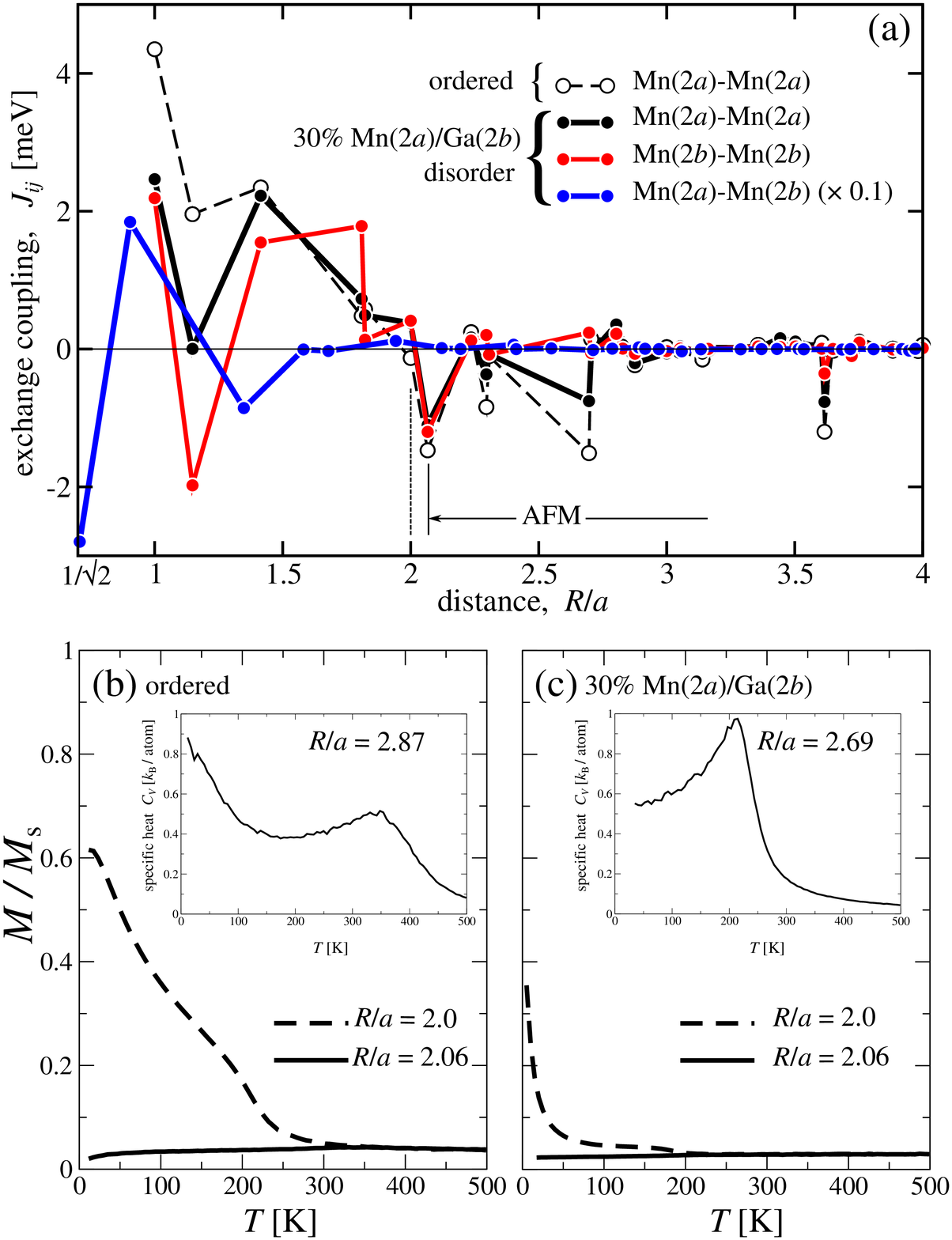}
\caption{\label{fig:theory-jxc} (a)~Isotropic exchange coupling constants $J_{ij}$ calculated as functions of distance $R$ (in
  the units of $a$) between $i$ and $j$ Mn sites for the fully ordered and partially
  disordered cases. (b) and (c)~represent the Monte-Carlo simulated
  ${M/M_{\rm S}}$ temperature dependencies of the Heisenberg model parametrized by
  the computed $J_{ij}$, in the fully ordered and partially disordered
  cases, respectively. Solid line corresponds to the minimal cluster size needed
  to set the AFM order, dashed - by one shall smaller cluster. The insets show the temperature dependency of the specific
  heat $C_{\rm V}$ (computed for the largest cluster size) which
  indicates the position of $T_{\rm N}$ by the local maximum.} 
\end{figure}

It is instructive to understand which mechanisms
are responsible for setting such ground-state modulation. Since the
relativistic effects do not affect the ground-state $\vec q$ vector substantially, in
the following we will calculate the  isotropic exchange coupling
constants $J$ using the real-space approach~\cite{LKAG87} implemented in
the SPR-KKR Green's function method~\cite{EKM11}.  In Fig.~\ref{fig:theory-jxc}\,a 
they are plotted as functions of distance between the interacting sites
$i$ and $j$. Here, we drop all interactions involving Pt and Ga atoms as insignificant, by leaving only
those between Mn atoms. In the fully ordered case, all nearest
Mn$(2a)$-Mn$(2a)$ interactions are parallel (${J>0}$), whereas the antiparallel 
ones (${J<0}$) are encountered by starting from ${R/a=2}$
(6-th shell within $(2a)$-sublattice). As it is shown by $M(T)/M_{\rm
  S}$ curves (Fig.~\ref{fig:theory-jxc}\,b) 
obtained by the Monte-Carlo simulation (ALPS package~\cite{BCE+11}) of the classical Heisenberg model
(${H=-\sum_{i>j}J_{ij}\,{\vec e}_i\cdot{\vec e}_j}$, where ${\vec e}_{i,j}$
are the unity vectors along the local magnetization directions on  $i$
and $j$ sites), the AFM order sets in by including all interactions at
least up to ${R/a\approx2.06}$ (7-th shell); accounting of the higher shells
does not affect the $M(T)$ behavior anymore.  Such a superposition of
the strong nearest parallel and the weaker long-range antiparallel exchange
interactions typically allows for the long-range spin-spiral order. Its direction (${\vec q}\parallel[001]$) follows from the
symmetry reasons: the  7-th shell, critical for setting up the AFM
order, contains 8 atoms at ${\vec R=(\pm a,0,\pm c)}$ and
${(0,\pm a,\pm c)}$, situated above and below the $ab$-plane of the
central atom. The corresponding Neel temperature can be estimated from
the peak of the magnetic specific heat $C_{\rm V}(T)$ computed for the
largest cluster size (${T_{\rm N}\approx350}$\,K at ${R/a\approx2.87}$,
see the inset in Fig.~\ref{fig:theory-jxc}\,b). This reasonably agrees
with experimental $M(T)$ slope change, well seen at about the same
temperature. 

The influence of chemical disorder can be estimated by means of the CPA
alloy theory~\cite{Sov67,Tay67}, implemented in the SPR-KKR method. By
assuming 30\,\% of Mn($2a$)/Ga($2b$) disorder, as it was identified in
experiment, modifies the exchange picture (Fig.~\ref{fig:theory-jxc}\,a). The exchange
interactions within Mn($2a$) sublattice, especially  the nearest-neighbor ones, become slightly weaker compared to the ordered
case. In addition, its statistical factor of ${(1-0.3)^2=0.49}$ might noticeably reduce the Neel temperature.  Altogether, this coupling
alone would lead to the similar magnetic order as in the fully ordered case.
Similar dependency is exhibited within the extra sublattice Mn($2b$),
except for the 2-nd shell, which shows a noticeable antiparallel coupling to 8 atoms at ${(\pm a/2,\pm a/2,\pm c/2)}$, which also leads to $\vec q\parallel$[001], though favoring a
shorter wavelength. However, since these interactions enter with the
low weight of ${0.3^2=0.09}$, they play a minor role for a final picture. 
The most important here is Mn$(2a)$-Mn$(2b)$ coupling, which has a
moderate statistical weight of ${0.3\cdot(1-0.3)=0.21}$, but also a huge
bare amplitude of the near-neighbor interactions (note that,
$J_{\text{Mn}(2a)-\text{Mn}(2b)}$ values shown in Fig.~\ref{fig:theory-jxc}\,a are downscaled by a factor of 10). In particular, within the $ab$-plane
there is a huge antiparallel interaction of  $-28$~meV with the 4 first-shell neighbors
at ${(\pm a/2,\pm a/2,0)}$, but also strong $+18$~meV
parallel out-of-plane interaction with two second-shell neighbors at ${(0,0,\pm c/2)}$. Alone, this
combination would lead to the vertical parallel coupling of the
neighboring collinear-AFM ordered 
$ab$-planes, however the last important $\sim-9$\,meV antiparallel coupling to 8 neighbors in a 3-rd shell at ${(\pm a,0,\pm
  c/2)}$, ${(0,\pm a,\pm c/2)}$ distorts this picture. 
Thus, to get a rough idea about the behavior of such complicated
magnetic system, we simplified the corresponding Heisenberg model, by
parametrizing it with effective interactions (i.e., by multiplying the bare
values with corresponding statistical weights). As it follows from Fig.~\ref{fig:theory-jxc}\,c, the minimal cluster size which sets the
AFM order remains the same, however, the $T_{\rm N}$
drops down to about 220\,K (see the inset). Since this
drives us away from the experimental result, it most probably indicates that the
effective parametrization of the Heisenberg model is inapplicable
(despite that the bare exchange interactions are correct) and
the adequate description might be achieved by treating the local effects
explicitly, which requires substantially larger cluster sizes.
